\documentclass[draft]{agujournal2019}
\usepackage{url} 
\usepackage{lineno}
\journalname{Journal of Advances in Modeling Earth Systems (JAMES)}

\begin{document}

%
%


\title{Data-driven super-parameterization using deep learning: Experimentation with multi-scale Lorenz 96 systems and transfer-learning}

%
%




\authors{Ashesh Chattopadhyay\affil{1}, Adam Subel\affil{1}, and Pedram Hassanzadeh\affil{1,2}}

\affiliation{1}{Department of Mechanical Engineering, Rice University, Houston, TX, USA}
\affiliation{2}{Department of Earth, Environmental and Planetary Sciences, Rice University, Houston, TX, USA}






\correspondingauthor{Pedram Hassanzadeh}{pedram@rice.edu}




\begin{keypoints}
\item{Data-driven super-parameterization (DD-SP), in which the equations of small-scale processes are integrated using deep learning, is proposed.}
\item{Tests with Lorenz systems find the cost-effective DD-SP having accuracy comparable to SP and outperforming data-driven parameterization.}
\item{Transfer-learning is shown to effectively improve the generalization of data-driven models to systems that are more chaotic.}
\end{keypoints}

%
%


\begin{abstract}
To make weather/climate modeling computationally affordable, small-scale processes are usually represented in terms of the large-scale, explicitly-resolved processes using physics-based or semi-empirical parameterization schemes. Another approach, computationally more demanding but often more accurate, is super-parameterization (SP), which involves integrating the equations of small-scale processes on high-resolution grids embedded within the low-resolution grids of large-scale processes. Recently, studies have used machine learning (ML) to develop data-driven parameterization (DD-P) schemes. Here, we propose a new approach, data-driven SP (DD-SP), in which the equations of the small-scale processes are integrated data-drivenly using ML methods such as recurrent neural networks. Employing multi-scale Lorenz 96 systems as testbed, we compare the cost and accuracy (in terms of both short-term prediction and long-term statistics) of parameterized low-resolution (LR), SP, DD-P, and DD-SP models. We show that with the same computational cost, DD-SP substantially outperforms LR, and is better than DD-P, particularly when scale separation is lacking. DD-SP is much cheaper than SP, yet its accuracy is the same in reproducing long-term statistics and often comparable in short-term forecasting. We also investigate generalization, finding that when models trained on data from one system are applied to a system with different forcing (e.g., more chaotic), the models often do not generalize, particularly when the short-term prediction accuracy is examined. But we show that transfer-learning, which involves re-training the data-driven model with a small amount of data from the new system, significantly improves generalization. Potential applications of DD-SP and transfer-learning in climate/weather modeling and the expected challenges are discussed.
\end{abstract}

%
%

%


%
%
%
%

\section{Introduction}
Some of the key components of the Earth system, such as atmospheric and oceanic turbulent circulations, involve multi-scale, multi-physics, chaotic processes. Because explicitly solving for all processes and scales is computationally prohibitive, in most of the current climate and weather models, which have, respectively, typical horizontal grid resolution of $O(100)$~km and $O(10)$~km, \textit{parameterization} schemes are used to represent small-scale, \textit{subgrid} processes (denoted with $Y$ or $Z$ hereafter) in terms of large-scale, \textit{resolved} processes (denoted with $X$ hereafter), e.g., as 
\begin{eqnarray}
Y=P(X),
\label{eq:P}
\end{eqnarray} 
where $P$ is a physics-based or semi-empirical function \cite{hourdin2017art,schneider2017climate,jeevanjee2017perspective,rasp2018deep,rasp2019online,palmer2019stochastic,chattopadhyay2019data}. However, despite much efforts, some of these often heuristic parameterization schemes, e.g., for clouds or gravity waves, have major shortcomings that result in persistent biases and large uncertainties, degrading weather forecasts and climate change projections \cite{alexander2010recent,sigmond2010influence,stevens2013climate,bony2015clouds,schneider2017climate,polichtchouk2018impact}.  

In the past two decades, an alternative approach called \textit{super-parameterization} (SP) has been explored, in which the governing equations of $Y$ are solved on a high-resolution grid embedded within each grid point of the low-resolution grid used for solving the governing equations of $X$ \cite{grabowski1999crcp,khairoutdinov2001cloud,majda2014new}. In practice, to reduce the computational cost, a simplified version of the governing equations of $Y$ (e.g., 2D rather than 3D) are solved; still, SP is expected to outperform parameterization because it is based on solving governing equations rather than heuristic approximations. Furthermore, the two-way coupling between the equations of $X$ and $Y$ in SP allows for accounting for transient responses of $Y$ to changes in $X$, which can be important when there is no distinct (spatial and/or temporal) scale separation between $X$ and $Y$. Such transient responses are missing from parameterizations such as Eq.~(\ref{eq:P}), which assume instantaneous quasi-equilibrium between $X$ and $Y$ \cite{yano2012convective,hassanzadeh2016linear,palmer2019stochastic}. The super-parameterized Community Atmospheric Model, SP-CAM \cite{khairoutdinov2001cloud}, has been shown to outperform the parameterized CAM in simulating the Madden-Julian Oscillation and some aspects of precipitation extremes  \cite{benedict2009structure,andersen2012moist,kooperman2018rainfall}. Super-parameterized E3SM \cite{hannah2020initial}, regional SP in a global circulation model (GCM) \cite{jansson2019regional}, SP in ocean modeling \cite{campin2011super}, and stochastic SP of geophysical turbulence \cite{grooms2013efficient} have been explored in recent years too and shown promising results. However, super-parameterized models are computationally much more demanding than parameterized models, which limits the applicability of SP.

In the past few years, data-driven modeling of subgrid processes using recent advances in machine learning (ML) has been explored in prototype chaotic dynamical systems and GCMs \cite{schneider2017earth,rasp2018deep,brenowitz2018prognostic,brenowitz2019spatially,o2018using,bolton2019applications,gagne2019machine,beucler2019achieving,yuval2020use}. In most of these studies, the aim is to learn a data-driven representation of $P$ in Eq.~(\ref{eq:P}) from observations and/or high-fidelity simulations. We refer to this general approach, which often employs feed-forward neural networks, as \textit{data-driven parameterization} (DD-P). While these studies have shown promising results, for practical use, issues such as ML method selection, stability of numerical solver-neural network coupled model, and generalization need to be further addressed. Generalization, the ability of a data-driven model to work accurately for a system with data distribution that is different from that of the training dataset, is particularly important, given the natural and anthropogenic non-stationarities in the climate system \cite{rasp2018deep}.

In this paper, we aim to further explore the applications of ML to improve the representation of subgrid processes. Specifically, the objective of this paper is two fold:
\begin{enumerate}
\item Introducing a novel, cost-effective framework: \textit{data-driven super-parameterization} (DD-SP),
\item Showing that \textit{transfer-learning} improves generalization.
\end{enumerate}
Regarding objective 1, recent work has shown that some types of recurrent neural networks (RNNs) can data-drivenly integrate differential equations of multi-scale chaotic systems accurately for hundreds of time steps \cite{chattopadhyay2019data}. Building on these results, we propose a DD-SP framework in which the equations of $Y$, which constitute the computationally expensive part of the SP framework, are integrated data-drivenly at high resolution using an RNN at low computational cost, while the equations of $X$ are integrated on a low-resolution grid using common numerical solvers. \textit{The overarching goal of this new approach is to achieve the accuracy of SP with the computational cost of physics-based or data-driven parameterization.}  

\begin{figure}[ht]
  \centering
  \includegraphics[height=\dimexpr \textheight - 4\baselineskip\relax]{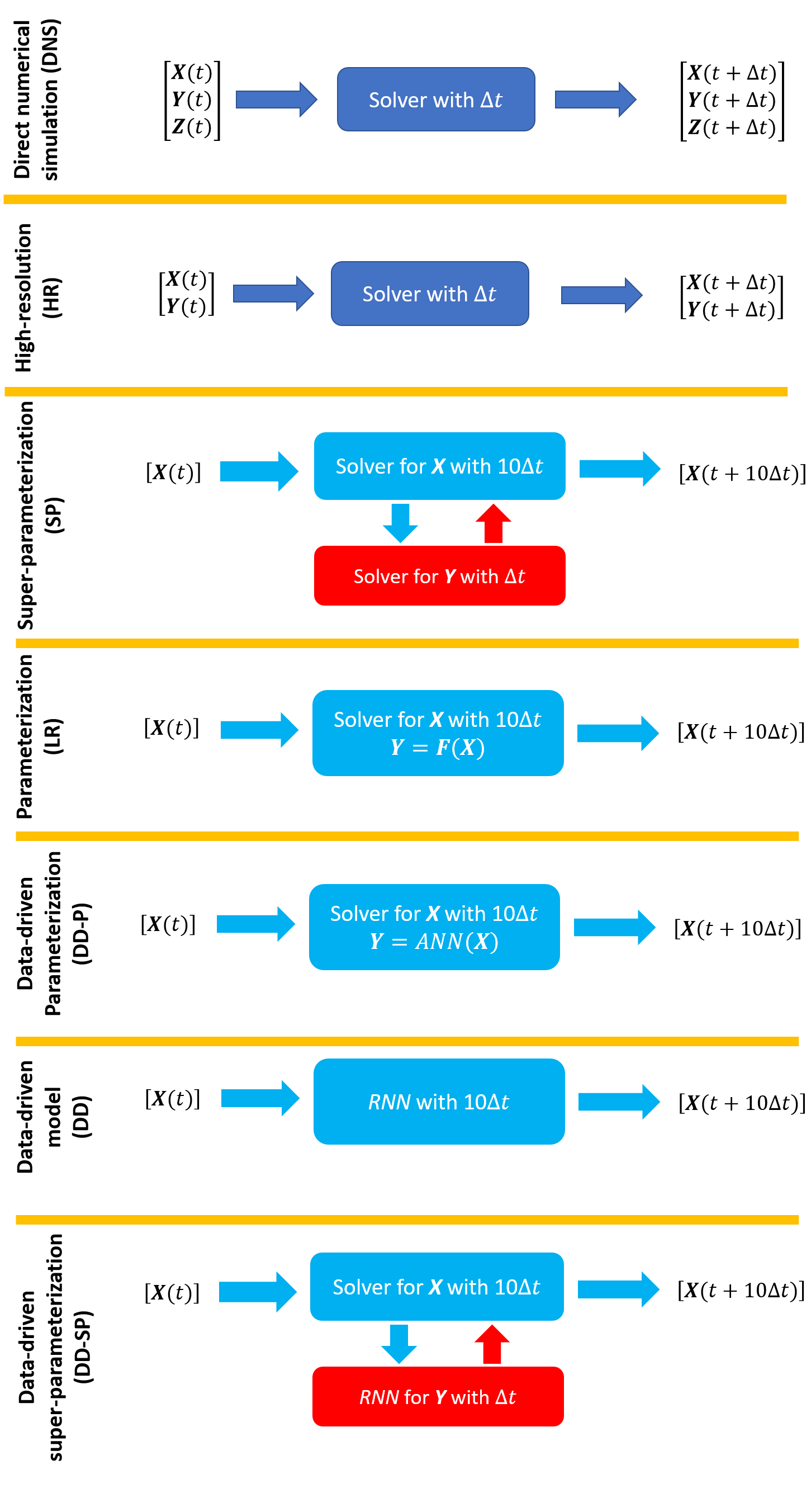}
  \caption{Schematic of the numerical and data-driven parameterized and super-parameterized models described in Section~\ref{method_section}. ``Solver'' refers to a common ODE/PDE numerical solver. ANN: Feed-forward artificial neural network; RNN: Recurrent neural network. }
\label{methods_cartoon}.
\end{figure}

To provide a proof-of-concept for DD-SP, here we use a chaotic Lorenz 96 system that has three sets of variables with different scales (in the order of decreasing scale: $X$, $Y$, and $Z$). For solving this system, we develop a number of models that mimic high-resolution (HR), parameterized low-resolution (LR), super-parameterized (SP), data-driven parameterized low-resolution (DD-P), data-driven super-parameterized (DD-SP), and fully data-driven (DD) climate/weather models (see Fig.~\ref{methods_cartoon}). We compare the computational cost and accuracy of these models for short-term spatio-temporal forecasting and reproducing the long-term statistics of the chaotic, multi-scale system. We also examine the effect of temporal scale separation between $X$ and $Y$ on the performance of DD-P and DD-SP.       

 
Regarding objective~2, we examine how DD-P, DD-SP, and DD models generalize when the forcing of the Lorenz system (and thus its chaoticity) increases. When generalization is poor, we show, for the first time to the best of our knowledge, that transfer-learning \cite{yosinski2014transferable}, which involves re-training the neural network with a small number of samples from the new system, improves generalization of data-driven models for subgrid processes. 

The paper is organized as follows: The multi-scale Lorenz 96 system and its numerical solution are described in Section~\ref{lorenz_eq}, followed by descriptions of the the numerical and data-driven parameterized and super-parameterized models in Section~\ref{method_section}. The performance of these models for short- and long-term spatio-temporal predictions are compared and the issue of generalization and application of transfer-learning are discussed in Section~\ref{results}. Potential applications of the proposed models and future work are discussed in Section~\ref{sec:discussion}.

\section{The Multi-scale Lorenz 96 System and Data}
\label{lorenz_eq}
To test the performance of the models mentioned above, we use a multi-scale Lorenz 96 system \cite{thornes2017use}:
\begin{eqnarray}
\label{lorenz}
\frac{dX_k}{dt}&=&X_{k-1}\left(X_{k+1}-X_{k-2}\right)-X_k+F-\frac{hc}{b}\Sigma_jY_{j,k} \label{eq:L1} \\ 
\frac{dY_{j,k}}{dt}&=&-cbY_{j+1,k}\left(Y_{j+2,k}-Y_{j-1,k}\right)-cY_{j,k}+  \frac{hc}{b}X_k -\frac{he}{d}\Sigma_iZ_{i,j,k} \label{eq:L2}\\
\frac{dZ_{i,j,k}}{dt}&=&edZ_{i-1,j,k}\left(Z_{i+1,j,k}-Z_{i-2,j,k}\right)- geZ_{i,j,k}+  \frac{he}{d}Y_{j,k} \label{eq:L3}
\end{eqnarray}
where $i, j, k = {1,2, \dots 8}$, i.e., there are $8$ equations for $X$, and $64$ and $512$ equations for  $Y$ and $Z$, respectively. This set of coupled nonlinear ODEs is a 3-tier extension of Lorenz's original model \cite{lorenz1996predictability} and has been used in several recent studies as a testbed for applications of ML to chaotic and weather/climate systems \cite{schneider2017earth,dueben2018challenges,mcdermott2019deep,watson2019applying,chattopadhyay2019data,gagne2019machine}.

In Eq.~(\ref{eq:L1}), $F$ is the large-scale forcing. Unless stated otherwise, $F=20$ is used, which makes the system highly chaotic. Coefficients $b,c,d,e,g,$ and $h$ determine the relative amplitudes and varibilities of $X$, $Y$, and $Z$. In this study, as discussed below, we use two sets of coefficients leading to Case 1, in which there is clear scale separation between $X$ and $Y$, and Case 2, in which there is little scale separation between these two variables.


\subsection{Case 1: Scale separation between $X$ and $Y$} \label{sec:Case1}
\begin{figure}[ht]
  \centering
  \includegraphics[width=1\textwidth]{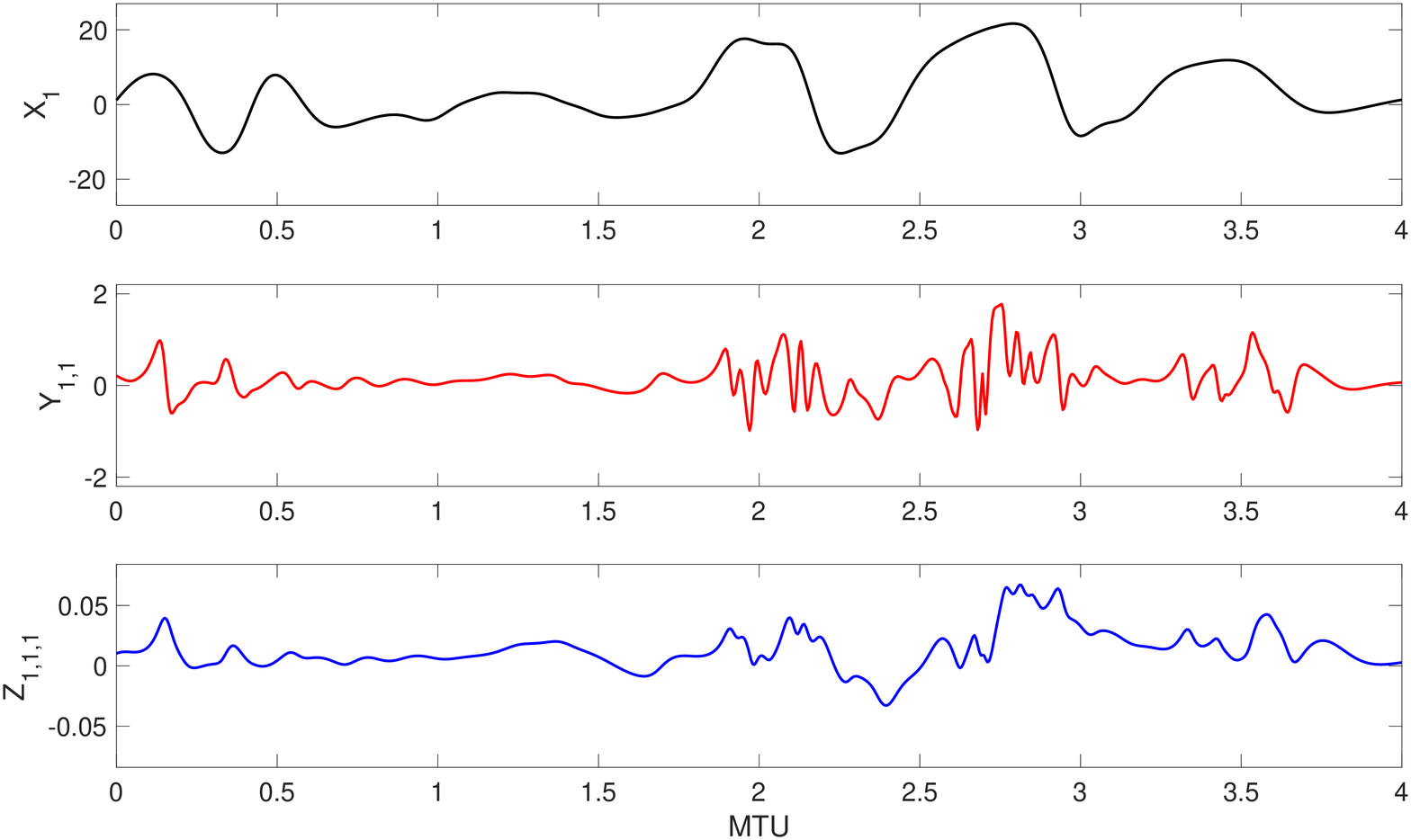}
  \caption{Examples of temporal evolution of $X_{1}$, $Y_{1,1}$, and $Z_{1,1,1}$ for Case 1. The time series show chaotic behavior in $X$, which has large amplitudes of $O(10)$ and low-frequency variability; $Y$, which has relatively small amplitudes of $O(1)$, high-frequency variability, and intermittency; $Z$, which has small amplitudes of $O(0.01)$ and high-frequency variability. There is temporal scale separation between $X$ and $Y$. The $x$-axis is in model time unit (MTU): $1$ MTU $=$ $200 \Delta t$, where $\Delta t$ is the time step of the numerical solver.}
\label{xyz_demo_uncoupled}.
\end{figure}

In this case, $b=c=e=d=g=10$ and $h=1$. Figure~\ref{xyz_demo_uncoupled} shows examples of the chaotic temporal evolution of $X$, $Y$, and $Z$ obtained from directly solving  Eqs.~(\ref{eq:L1})-(\ref{eq:L3}) using a $4$th-order Runge-Kutta (RK4) scheme (see Section~\ref{sec:DNS}). These examples demonstrate that $X$ has large amplitudes and slow variability; $Y$ has relatively small amplitudes, high-frequency variability, and intermittency; $Z$ has small amplitudes and high-frequency variability. In this case, there is (temporal) scale separation between $X$ and $Y$, which justifies assuming instantaneous quasi-equilibrium between $X$ and $Y$, and using parameterizations of the type of Eq.~(\ref{eq:P}) with physics-based or data-driven $P$ \cite{arnold2013stochastic,hassanzadeh2016linear,khodkar2019reduced}.         


\subsection{Case 2: No scale separation between $X$ and $Y$} \label{sec:Case2}
\begin{figure}[ht]
  \centering
  \includegraphics[width=1\textwidth]{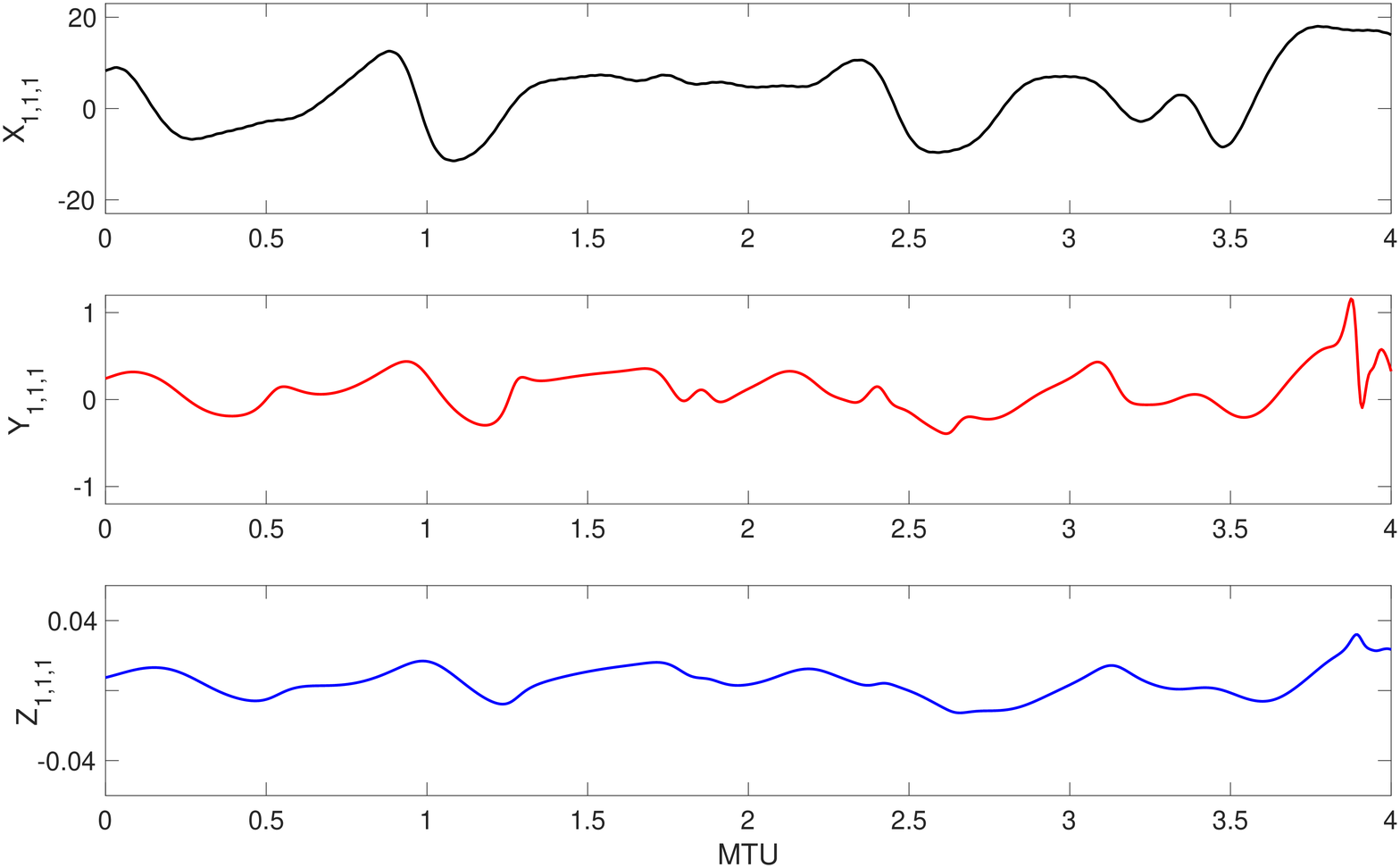}
  \caption{Same as Fig.~\ref{xyz_demo_uncoupled} but for Case 2. The time series show chaotic behavior in $X$, $Y$, and $Z$, which like the variables of Case 1, have different amplitudes of $O(10)$, $O(1)$, and $O(0.01)$, respectively. However, unlike the variables of Case 1, $X$ and $Y$ (and $Z$) do not have distinct temporal time scale separation.}
\label{xyz_demo_coupled}.
\end{figure}

In this case, $b=e=d=g=10$, $c=8$, and $h=0.5$. Figure~\ref{xyz_demo_coupled} shows examples of the chaotic temporal evolution of $X$, $Y$, and $Z$. In Case 2, like in Case 1, these variables have different amplitudes. However, unlike in Case 1, $X$ and $Y$ (and $Z$) have similar time scales and there is no distinct temporal scale separation. Therefore, in Case 2, one \textit{cannot} assume quasi-equilibrium between $X$ and $Y$, i.e., $Y$ can be determined only from $X$ and the transient response of $Y$ becomes important. Lack of scale separation makes parameterizations of the type of Eq.~(\ref{eq:P}) vulnerable to inaccuracies \cite{palmer2019stochastic,yano2012convective}. Note that many processes in the weather/climate system lack clear spatio-temporal scale separations \cite{gross2018physics}.               


\section{The Numerical and Data-driven (Super-) Parameterized Models}
\label{method_section}
Below we describe the various numerical and data-driven parameterized or super-parameterized models that are compared in Section~\ref{results}. These models are schematically shown in Fig.~\ref{methods_cartoon}. Their computational costs are summarized in Table~\ref{table_cost}. Note that in designing the models and assessing their accuracy and cost, we assume that only predicting the large-scale variable $X$ at time scales longer than $10\Delta t$ is of interest; we might solve for the small-scale/high-frequency variables $Y$ and $Z$, but only because of their effects on $X$, and these variables themselves are not of interest.  

\begin{table}
  \centering
 \caption{Normalized computational cost of each model. Cost is evaluated as the number of equations that are needed to be solved numerically to compute $X(t+10 \Delta t)$ from $X(t)$. Costs are normalized such that the cost of DNS is $1000$. Computational cost of neural networks in DD, DD-P, and DD-SP is assumed $0$.}  
\centering

\begin{tabular}{ | c | c | c | c | c | c | c | c |}
\hline
  Model & Direct  & High- & Super-              & Data-driven  & Fully       & Data-driven   & Low- \\
        &  numerical        & resolution                & parameterized       & super-           & data-driven & parameterized & resolution \\
        & simulation & & & parameterized  & & & \\
        &  (DNS)             & (HR)            & (SP)                & (DD-SP)      & (DD)        & (DD-P)        & (LR) \\ 
\hline 
 Cost & 1000 & 123.3 & 111.0 & 1.4 & 0 & 1.4 & 1.4 \\  
\hline
\end{tabular}
 \label{table_cost}
\end{table}


\subsection{Direct Numerical Simulation (DNS)}\label{sec:DNS}
In DNS, Eqs.~(\ref{eq:L1})-(\ref{eq:L3}) are numerically integrated using a RK4 scheme with time step $\Delta t=0.005$. This time step is dictated by the numerical accuracy/stability requirements imposed by high-frequency variables $Y$ and $Z$. Here, the terminology of ``DNS'' is used for convenience to refer a model in which all scales are resolved and no equation is simplified or ignored. Data from DNS are taken as the ``truth'' and the accuracy of each models is compared with respect to DNS. DNS data are used for the training and validation of data-driven models.

\subsection{The High-resolution (HR) Model}
In HR, Eqs.~(\ref{eq:L1}) and (\ref{eq:L2}) are integrated numerically using a RK4 scheme with time step $\Delta t$. Variable $Z$ (Eq.~(\ref{eq:L3})) and the last term in Eq.~(\ref{eq:L2}) are ignored. Note that $Z$ could be parameterized in terms of $Y$ (as for example, done in \citeA{thornes2017use}), but that would lead to little improvements in the model. The HR model is computationally cheaper than DNS (by a factor of $\sim 8$), but still represents the class of models that are currently computationally too expensive to use in practice.      

\subsection{The Low-resolution (LR) Model}
In this model, evolution of $X$ is obtained from numerical integration of an equation that is similar to Eq.~(\ref{eq:L1}), but in which the effects of $Y$ on $X$ are parameterized using function $U_p$ 
\begin{eqnarray}
\frac{dX_k}{dt}=X_{k-1}\left(X_{k+1}-X_{k-2}\right)-X_k+F-U_p(X_k) \label{eq:LR1}
\end{eqnarray}
Following \citeA{thornes2017use}, we use the following stochastic parameterization scheme: 
\begin{eqnarray}
U_p(X)&=&U_{det}(X)+e(t)\\
U_{det}(X)&=&a_0+a_1X+a_2X^2+a_3X^3+a_4X^4 \label{eq:U} \\
e(t)&=&\phi  e(t-\Delta t)+\sigma_e(1-\phi^2)^{\frac{1}{2}}z(t)
\end{eqnarray} 
where $\phi$ is the lag-1 auto-correlation, $\sigma_e$ is the standard deviation of the stochastic tendency, and $z(t)$ is white noise with unit standard deviation; $z(t) \sim \mathcal{N}(0,1)$. The coefficients of Eq.~(\ref{eq:U}) are determined using a $4$th order polynomial fit. 

Equation~(\ref{eq:LR1}) is solved numerically with a RK4 scheme with time step $10 \Delta t$. Note that in this ``low resolution'' model, we are not solving any equation for $Y$ (and ignore $Z$), which allows us taking larger time steps in the RK4 solver. The LR model is computationally much cheaper than the HR model (by a factor of $\sim 88$; see Table~\ref{table_cost}), and represents the class of models that are most commonly used in practice, i.e., low resolution with semi-empirical or physics-based parameterization.       



\subsection{The Super-parameterized (SP) Model}\label{sec:SP}
As summarized in Fig.~\ref{methods_cartoon}, in SP:
\begin{enumerate}
    \item Eq.~(\ref{eq:L1}) is numerically integrated using a RK4 scheme with time step $10 \Delta t$, e.g., to find $X(t)$ from $X(t-10\Delta t)$ (and using $Y(t-10\Delta t)$ and $Y(t)$).
    \item $X(t)$ is fed into Eq.~(\ref{eq:L2}), which is numerically integrated using a RK4 scheme with time step $\Delta t$ to find $Y(t+10\Delta t)$ from $Y(t)$. Variable $Z$ (Eq.~(\ref{eq:L3})) and the last term in Eq.~(\ref{eq:L2}) are ignored.
    \item The computed $Y(t+10\Delta t)$ is then used in Step~1 to find $X(t+10\Delta t)$ and the cycle continues.
\end{enumerate}
This setup follows the same SP philosophy for dealing with multi-scale systems used in super-parameterized GCMs such as SP-CAM. However, in such models, a simplified version of the equations of small-scale processes (e.g., 2D rather than 3D) are solved to reduce the computational cost, while here, given the simplicity of the Lorenz system, Eq.~(\ref{eq:L2}) is solved in Step~2 without any simplification other than ignoring the term that involves $Z$. As a result, our SP model is just slightly cheaper than the HR model. Still, this setup is sufficient for our objective, which is to introduce and provide a proof-of-concept for the DD-SP model (Section~\ref{sec:DD-SP}), which is built on this SP model.

 

\subsection{The Data-driven Parameterized (DD-P) Model}
 This model is similar to LR, except that in Eq.~(\ref{eq:LR1}), a data-driven representation of $U_p$ is obtained from a feed-forward fully connected artificial neural network (ANN). Details of the ANN used here are in~\ref{sec:ANN_append}. The ANN is trained with $10^6$ sequential pairs of $X$ and $Y$ that are sampled every $\Delta t$ from the DNS data. The computational cost of DD-P is the same as LR, and it represents the class of models developed in some recent studies in which data-driven representations of subgrid processes in the atmosphere and ocean are learned using ML methods such as random forests, convolutional neural networks (CNNs), or ANNs \cite{rasp2018deep,o2018using,brenowitz2019spatially,bolton2019applications,salehipour2019deep,yuval2020use}.

\subsection{The Data-driven Super-parameterized (DD-SP) Model} 
The DD-SP model is the same as the SP model (Section~\ref{sec:SP}), except that in Step~2, Eq.~(\ref{eq:L2}) is integrated data-drivenly using an RNN rather than using a numerical solver. The RNN we use here is a gated recurrent unit (GRU); see \ref{append_GRU} for details. Similar to the ANN, the GRU is trained with $10^6$ sequential pairs of $X$ and $Y$ that are sampled every $\Delta t$ from the DNS data. Once trained, with an input of $X(t)$ and $Y(t)$, the GRU predicts the spatio-temporal evolution of $Y$ and computes $Y(t+\Delta t), Y(t+2\Delta t), \dots Y(t+10 \Delta t)$. The last value is fed into Eq.~(\ref{eq:L1}), which is then numerically integrated to find $X(t+10\Delta t$) (Step~3), and the cycle continues.    

Because of the data-driven rather than numerical integration of Eq.~(\ref{eq:L2}), which requires high resolutions, the DD-SP model is computationally much cheaper than the SP model (by a factor of $\sim 88$), and in fact, has the same cost as that of the LR or DD-P model (with the caveat that here we are assuming that ANNs and GRUs have negligible computational costs compared to numerical solvers such as RK4; see Section~\ref{sec:metric}).     


\subsection{The Fully Data-driven (DD) Model} \label{sec:DD-SP}
Although the purpose of this paper is to explore data-driven representation of subgrid processes in numerical models, we have also investigated the performance of a fully DD model. In this model, following \citeA{chattopadhyay2019data}, an RNN is trained on $X$, which is then used to predict the spatio-temporal evolution of $X$ from an initial condition. The RNN we use here is a GRU (see \ref{append_GRU_DD}), trained on $10^6$ sequential values of $X(t)$ sampled at every $10 \Delta t$ from the DNS data.       


\subsection{Metrics for Performance Comparison}\label{sec:metric}
To compare the performance of the proposed models presented in Section \ref{method_section}, we measure their accuracy and computational cost. To find the accuracy of short-term forecasting, we first compute the relative error
\begin{eqnarray}
e(t)=\frac{||X_{pred}(t)-X_{DNS}(t)||}{\frac{1}{(N+1)}\sum_{\tau=0}^{\tau=N \Delta t}||X_{DNS}(\tau)||}
\label{eq:e}
\end{eqnarray}
averaged over $100$ randomly chosen initial conditions, each of them outside the training set. $||.||$ is the $L_2$ norm of a vector; $N$ is typically a large integer value ($\approx 2000$). We report prediction horizon, in terms of $\Delta t$, defined as when the running mean of $e(t)$ with a window of $4\Delta t$ reaches $0.3$. We also report the average of $e(t)$ over the first $100$ or $200 \Delta t$. For each initial condition, the prediction horizon is first calculated and then the mean over all $100$ (randomly chosen) initial conditions is evaluated. 

To evaluate how well a model reproduces the long-term statistics of the system, we compare the probability density function (PDF) of the data produced from a long integration by the model with the PDF of the DNS data. We perform a two-sample Kolmogorov-Smirnov (KS) test \cite{massey1951kolmogorov} between the samples from the model data and the DNS data and report the probability ($p$) of the samples coming from the same distribution. A high (low) probability indicates that the PDF of the model data is similar (not similar) to the PDF of the DNS data. 


For each model, computational cost is evaluated as the number of equations that are needed to be solved numerically using the RK4 method to compute $X(t+10 \Delta t)$ from $X(t)$. Costs are normalized such that the cost of DNS is $1000$. We assume the parameterization in the LR model, and the ANN or GRU in the DD-P, DD-SP, and DD models to have negligible computational costs once trained. In reality, there is a computational cost associated with running an ANN or GRU, as well as associated with data transfer between the CPUs and GPUs. However, to avoid the complex issues related to implementation and algorithm/hardware used for numerical solvers versus neural networks, we choose not to use more sophisticated measures of cost such as the number of floating point operations or wall-clock time in this study. The normalized computational cost of each of the models is presented in Table~\ref{table_cost}.         


  
\section{Results}
\label{results}
\subsection{Short-term Spatio-temporal Forecasting} \label{sec:shorterm}
\subsubsection{Precise initial conditions}\label{sec:exactIC}
\begin{figure}[ht]
  \centering
\includegraphics[height=\dimexpr \textheight - 6\baselineskip\relax]{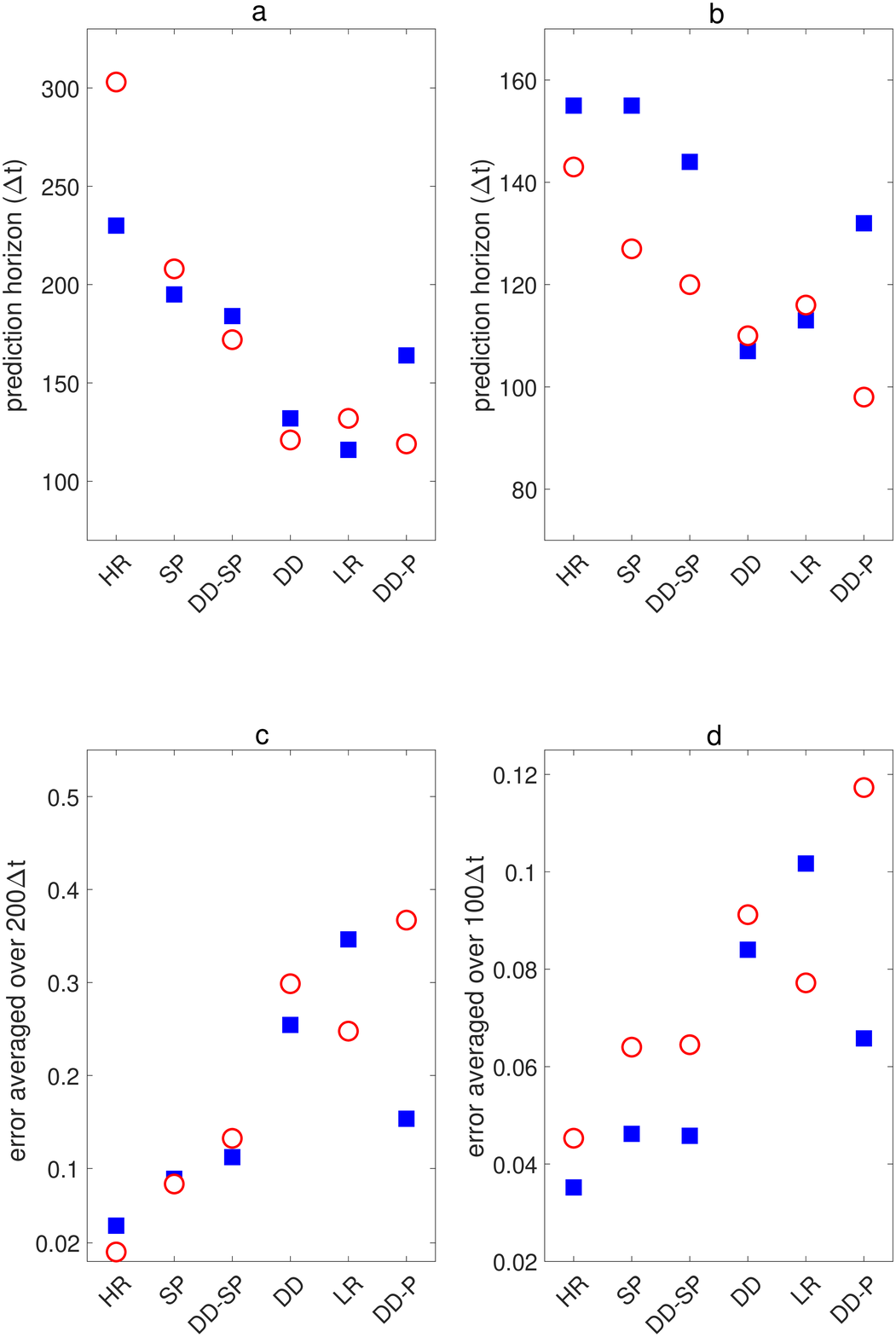}
  \caption{Comparison of the short-term prediction accuracy of all models. Prediction horizon averaged over predictions from 100 randomly chosen initial conditions of $X$ that are (a) precise (noise free) and (b) noisy. Relative error (Eq.~(\ref{eq:e})) averaged over predictions from 100 randomly chosen initial conditions of $X$ that are (c) precise (noise free) and (d) noisy. Blue squares: Case 1 (Section~\ref{sec:Case1}). Red circles: Case 2 (Section~\ref{sec:Case2}). Note that for noisy initial conditions, the prediction horizons are shorter, leading to different ranges of $y$-axis in (a) and (b), and using a shorter interval for averaging the error in (d) compared to (c).}
\label{all_cases_noisy}.
\end{figure}

The short-term prediction accuracy of all models is compared in Fig.~\ref{all_cases_noisy}. In (a) and (c), the prediction horizon (in terms of $\Delta t$) and the relative error $e(t)$ averaged over the first $200\Delta t$ for predictions starting form precise (noise free) initial conditions for both Cases 1 and 2 are shown. Averaged over 100 randomly chosen initial conditions, among the purely numerical models, for both cases, HR has the best performance, followed by SP, and then LR. For example, for Case 1, the prediction horizons of HR and SP are, respectively, $\approx 2$ and $\approx 1.7$ times longer than that of LR, which has the prediction horizon of $\approx 116\Delta t$. The better performance of the HR and SP models comes with a higher computational cost, about $88$ times that of LR (Table~\ref{table_cost}). 

Next, we examine the performance of the DD-P and DD-SP models, which as a reminder, have the same low computational cost as the LR model. For Case~1, both DD-P and DD-SP substantially outperform LR, respectively, by factors of $1.41$ and $1.59$ in prediction horizon (Fig.~\ref{all_cases_noisy}(a)), and by factors of $0.44$ and $0.32$ in averaged error (Fig.~\ref{all_cases_noisy}(c)). The DD-SP model outperforms DD-P, although by a relatively small margin (factors of $1.1$ in prediction horizon and $0.73$ in averaged error). Note that in this case, DD-SP has prediction horizon and averaged error close to those of SP (e.g., $192\Delta t$ versus $184\Delta t$), while being around $88$ times cheaper computationally. 

For Case~2, DD-P is even worse than LR (see Fig.~\ref{all_cases_noisy}(a) and (c)), while DD-SP again markedly outperforms LR, by factors of $1.3$ in prediction horizon and $0.53$ in averaged error. The inability of DD-P to provide any improvement over LR is not surprising, as the absence of scale separation invalidates the quasi-equilibrium assumption needed for parameterization of the type of Eq.~(\ref{eq:P}). Even though the super-parameterization framework (whether SP or DD-SP) has difficulties in dealing with systems that lack scale separation too (evident in the larger gap between HR and SP for Case~2 compared to Case~1 in panel (a)), the DD-SP model still provides a substantial improvement over LR. In this case, DD-SP is not as close to SP as in Case 1, e.g., the prediction horizon of SP is $208 \Delta t$ while that of DD-SP is $172 \Delta t$; however, as discussed in the next section, if the initial conditions are noisy rather than precise, the difference between the accuracy of SP and DD-SP largely vanishes. 

Note that Fig.~\ref{all_cases_noisy} also shows the accuracy of fully data-driven prediction (DD), which is often comparable to that of LR. Although the focus of this paper is on parameterization and super-parameterization, we include the DD results mainly for examining generalization, which is discussed in Section~\ref{sec:gen}.  

\begin{figure}[ht]
  \centering
  \includegraphics[width=1\textwidth]{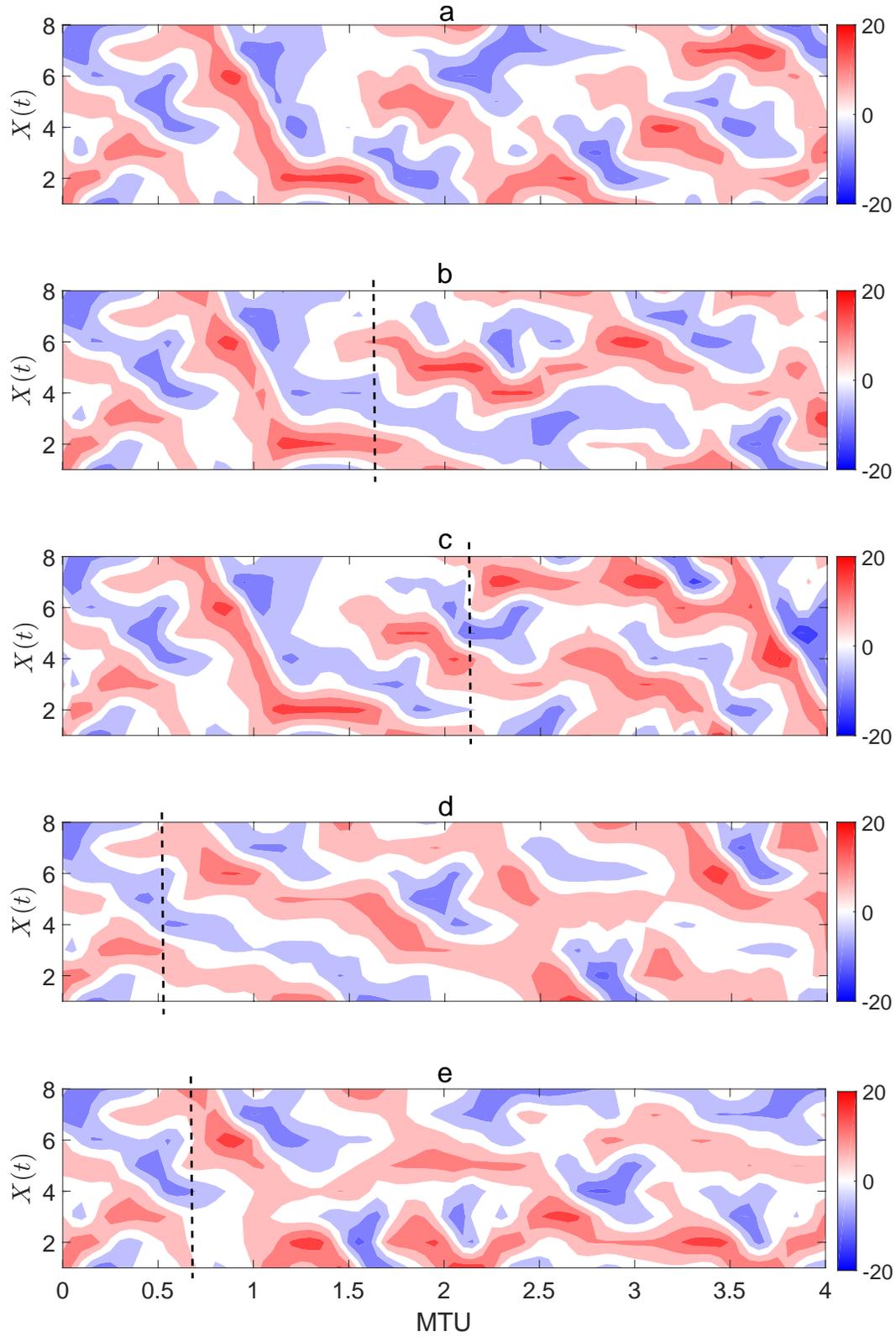}
  \caption{Examples of short-term prediction from the same initial condition for Case 1. (a) DNS (truth). (b) SP. (c) DD-SP. (d) LR. (e) DD-P. The dashed line on each panel shows the prediction horizon. $1$ MTU is $200 \Delta t$. For this case, the $e$-folding decorrelation time of the first principal component time series of $X$ is $\tau_{PC1}=28 \Delta t$.}
\label{contour_uncoupled}.
\end{figure}

Figure~\ref{all_cases_noisy} is based on relative error averaged over 100 random initial conditions and provides a fitting measure for a quantitative comparison of different models. Still, to give the readers another view of the skills of these models, Figs.~\ref{contour_uncoupled} and \ref{contour_uncoupled} show examples of the spatio-temporal evolution of $X$ from DNS and predicted using SP, DD-SP, LR, and DD-P for Cases 1 and 2, respectively. For each case, we intentionally pick an initial condition for which DD-SP outperforms SP, to further demonstrate the capabilities of DD-SP.

\begin{figure}[ht]
  \centering
  \includegraphics[width=1\textwidth]{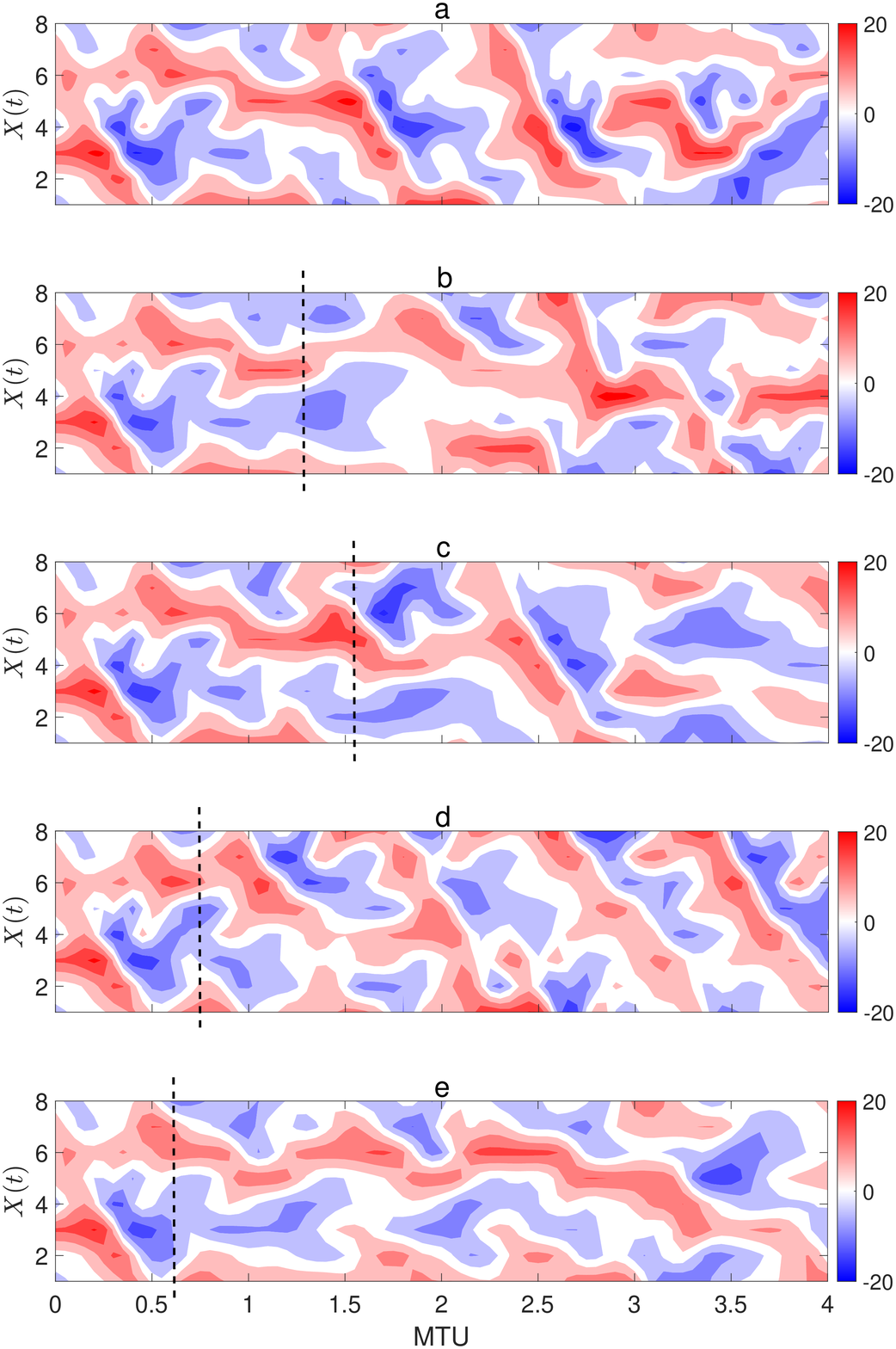}
  \caption{Same as Fig.~\ref{contour_uncoupled} but for Case 2. For this case, the $e$-folding decorrelation time of the first principal component time series of $X$ is $\tau_{PC1}=32 \Delta t$.}
\label{contour_coupled}.
\end{figure}
 
 \subsubsection{Noisy initial conditions}
To compare the performance of different models for the more realistic condition that the initial measurements of $X$ are not precise, we repeat the experiments of Section~\ref{sec:exactIC} but with noisy initial conditions. Random perturbations drawn from $\mathcal{N}(0,0.05 \sigma_X)$, where $\sigma_X$ is the standard deviation of $X$, are added to each initial condition and prediction is conducted. This is repeated for each initial condition $20$ times. As expected the prediction horizons of all models decrease, but much more for more accurate models such as HR and SP (Fig.~\ref{all_cases_noisy}(b) and (d)). While the ranking of the models (in terms of performance) does not change compared to what we find with precise initial conditions, the advantages of DD-SP over other models become clearer. For example, in comparison to computationally cheap models, for both Case 1 (2), the averaged error of DD-SP is $0.45$ $(0.83)$ of that of LR and $0.69$ $(0.55)$ of the averaged error of DD-P. In comparison with SP, which is $88$ times costlier, DD-SP has a quite similar averaged error and just slightly lower prediction horizon (by a factor of around $\approx 0.93$) for both Cases 1 and 2.

 \subsection{Reproducing Long-term Statistics} 
 \begin{table}
  \centering
 
 \caption{Two-sample KS tests between the PDF generated from a long-term integration of each model and the PDF of the DNS data for Cases 1 and 2. $p$ is the probability that the samples are taken from the same distribution. High (low) $p$ shows good (poor) agreement between the model-generated PDF and the PDF of the DNS data.}  
\centering
\begin{tabular}{ | c | c | c | c | c | c | c |}
\hline
  Model & HR & SP & DD-SP & DD & DD-P & LR \\
  \hline
  $p$: Case 1 & 1.00 & 1.00 & 0.99 & 0.99 & 0.99 & 0.90  \\ \hline
  $p$: Case 2 & 1.00 & 1.00 & 0.99 & 0.99 & 0.75 & 0.71  \\
  \hline 
\end{tabular}
 \label{KS_table_cases}
\end{table}

\begin{figure}[ht]
  \centering
  \includegraphics[width=1\textwidth]{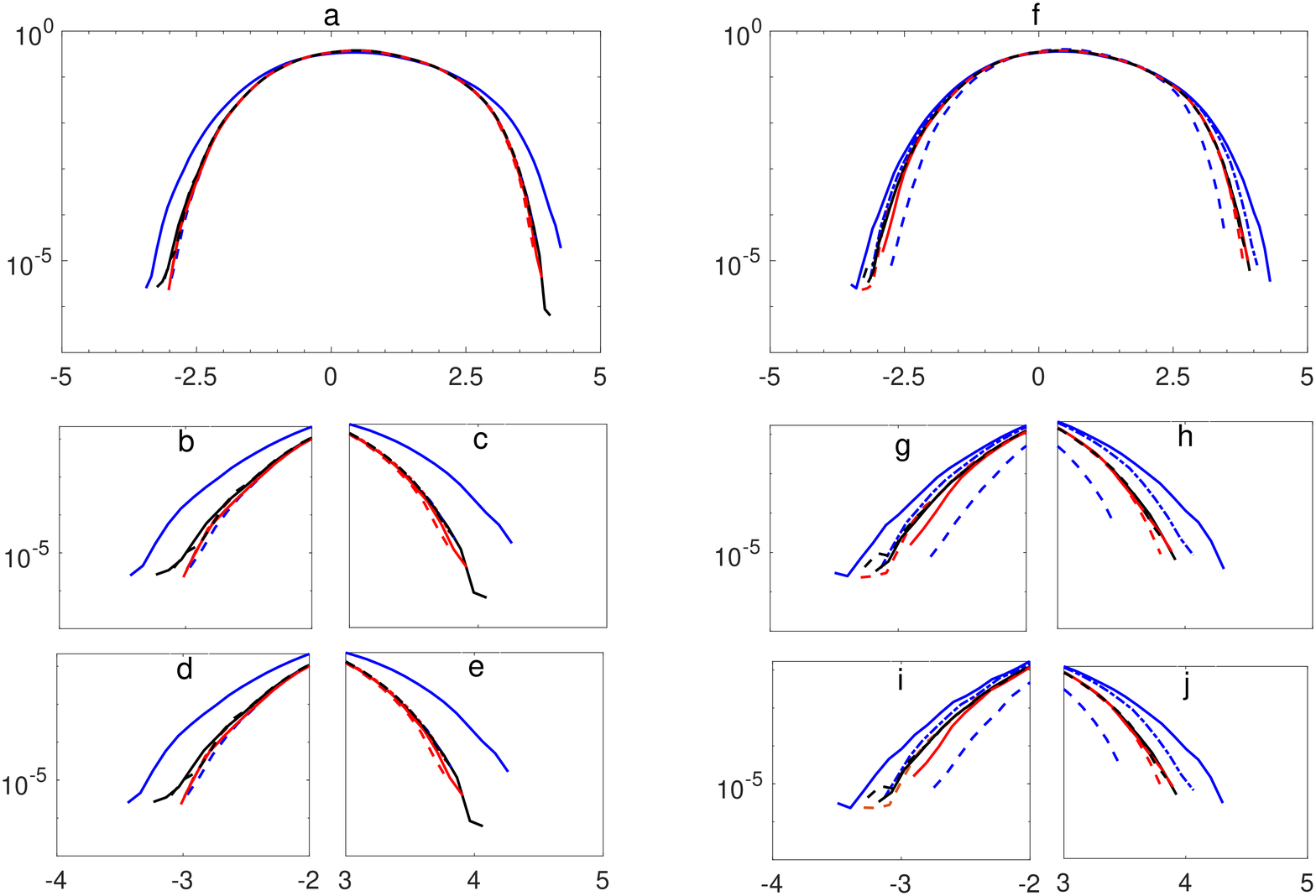}
  \caption{Comparison of the PDFs of data produced by all models. The PDF of DNS is computed using a dataset of length $10^8 \Delta t$ with sampling interval of $10 \Delta t$ and taken as the truth. The PDFs of the rest of the models are computed using datasets of length $35 \times 10^6 \Delta t$ with sampling interval of $10 \Delta t$. The kernel density estimator method \cite{epanechnikov1969non} is used to obtain the PDFs. (a)-(e): Case 1. (f)-(j): Case 2. (b)-(c) show the tails of (a) while (d)-(e) show the tails of Case 1 from another independent dataset of the same length. (g)-(j): Same as (b)-(e) but for Case 2. The similarities between the tails of PDFs of two independent long-term integrations show their robustness, allowing to draw conclusions about the ability of these models in simulating the statistics of the rare events. Lines show DNS (black solid), HR (black dashed), SP (red solid), DD-SP (red dashed), LR (blue solid), DD-P (blue dashed), and DD (blue dot-dashed).}
\label{pdf}.
\end{figure}

The short-term forecasting results of Section~\ref{sec:shorterm} are most relevant to weather forecasting. Below, we examine the ability of each model in reproducing the long-term statistics of the system, i.e., in simulating the system's climate. In Fig.~\ref{pdf}, the PDFs of the data generated from a long integration of each model are compared with the PDF of the DNS data. We particularly examine the tails of the PDFs, because accurately reproducing the statistics of the rare events of a system is important, e.g., to study the climate extremes. Note that ``long integration'' refers to integrations that continue well after the predicted trajectories deviate from the DNS trajectories (note that we find all models that used ANN or GRU to be numerically stable for as long as the integration is continued; see \ref{append_GRU} and Section~\ref{sec:discussion}). 

Figures~\ref{pdf}(a)-(e) and two-sample KS tests in Table~\ref{KS_table_cases} show that for Case~1, with the exception of LR, all models accurately reproduce the long-term statistics of the system, even at the tails. However, for Case 2, Figs.~\ref{pdf}(f)-(j) and the KS tests show that the PDFs produced by LR or DD-P do not match the PDF of DNS, and in particular, there are large deviations at the tails. The DD model captures the PDF well, although at the right tail (Figs.~\ref{pdf}(h) and (j)), the match is not as good as that of DD-SP. These results show the clear advantage of DD-SP as the best-performing low-cost model in reproducing the long-term statistics of both Cases 1 and 2, even at the tails of the PDFs.

\subsection{Generalization to Systems with Higher $F$: A transfer-learning Approach }
\label{sec:gen}
Generalization (i.e., extrapolation) of ML algorithms to new data distributions has remained a challenge and a source of concern in many applications \cite{sugiyama2012machine}. In climate modeling, the natural and anthropogenic variabilities cause non-stationarities, which can limit the applicability of a data-driven model that is trained with a dataset that contains a large amount of data but from a small part of the non-stationary distribution. For example, \citeA{rasp2018deep} reported that an ANN-based DD-P model for moist convection did not generalize when the global sea-surface temperature increased by $4$~K. In a more recent study by \citeA{schergeneralization}, the authors report the challenges for generalization of an ANN-based fully data-driven model to an increase in forcing $F$ for a much simpler Lorenz system (Lorenz 63 system). In this section, we explore the generalization of DD-SP, DD-P, and DD when forcing $F$ is increased from $20$ to $24$, thus increasing the chaoticity of the system.

\begin{figure}[ht]
  \centering
  \includegraphics[width=1\textwidth]{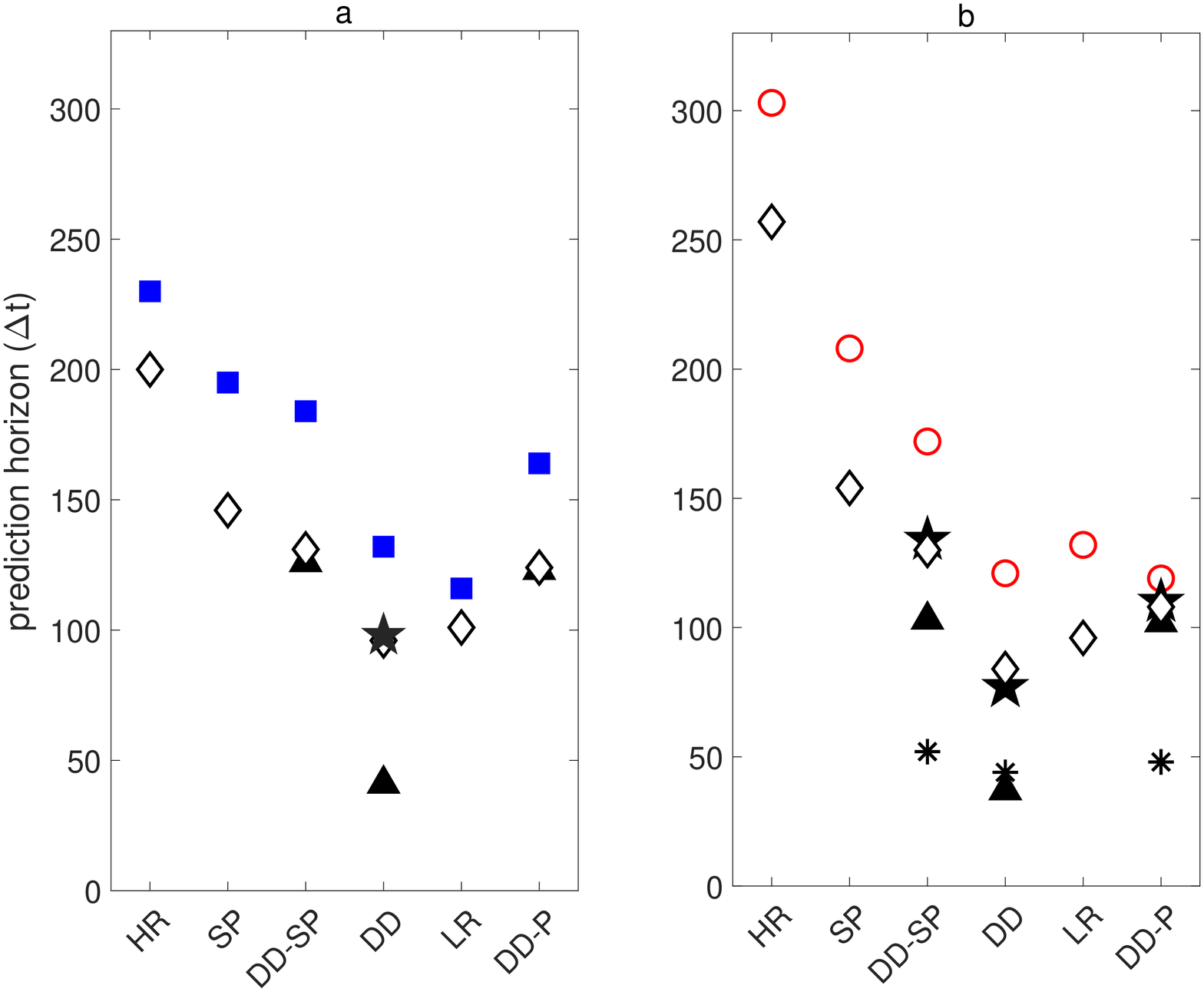}
  \caption{Generalization performance with and without transfer-learning in terms of short-term forecasting. (a) Case~1. (b) Case~2. Blue squares: training and testing with $F=20$ (same as in Fig.~\ref{all_cases_noisy}). For all other symbols, the models is tested for a system with $F=24$. Black diamonds: the ANN or GRU is trained on data from $F=24$. Black triangles: the ANN or GRU is trained on data from $F=20$. Black pentagram: the ANN or GRU is trained with transfer-learning using $10$K samples from $F=24$ and weights initialized with $W_{F=20}$. Black asterisks: ANN or GRU is trained with $10$K samples but with random initialization of weights (no transfer-learning).}
\label{TL}.
\end{figure}

 
The blue squares and black diamonds in Fig.~\ref{TL}(a) show the short-term prediction accuracy of all models for Case~1 when the training and testing of the models involving a data-driven component (DD-P, DD-SP, and DD) are conducted on data from the same system (i.e., same $F$). As expected, increasing $F$ leads to a decrease in prediction horizon for both numerical and data-driven models (this is also true for Case~2, see Fig.~\ref{TL}(b)). For Case~1, when DD and the ANN or GRU in DD-P and DD-SP are trained on data from the system with $F=20$ but the model is tested on a system with $F=24$ (black triangles), we find DD-P and DD-SP to generalize (black diamonds and triangles overlap), while DD fails to generalize (i.e., there is a gap between the diamond and triangle). To be clear, in these experiments, for DD-P and DD-SP, the ANN or GRU are trained on data from $F=20$, but the equations for $X$ are integrated numerically with $F=24$. Thus we are examining the generalization of data-driven subgrid-scale modeling.

\begin{table}
  \centering
\caption{Two-sample KS tests between the PDF generated from long-term integration of each model and the PDF of the DNS data for Case 2 without transfer-learning (trained on $F=20$ and tested on $F=24$) and with transfer-learning (re-training on $F=24$ with $1\%$ of original training size). $p$ is the probability that the samples are taken from the same distribution. High (low) $p$ shows good (poor) agreement between the model-generated PDF and DNS PDF.}  
\centering
\begin{tabular}{ | c | c | c | c | c | c | c |}
\hline
  Model & HR & SP & DD-SP & DD & DD-P & LR \\
  \hline
    $p$: without transfer-learning  & 1.00 & 1.00 & 0.99 & 0.99 & 0.91 & 0.91  \\
\hline
  $p$: with transfer-learning & 1.00 & 0.99 & 0.99 & 0.99 & 0.92 & 0.90  \\ 
  \hline 
\end{tabular}
 \label{KS_table_TL}
\end{table}

\begin{figure}[ht]
  \centering
  \includegraphics[width=1\textwidth]{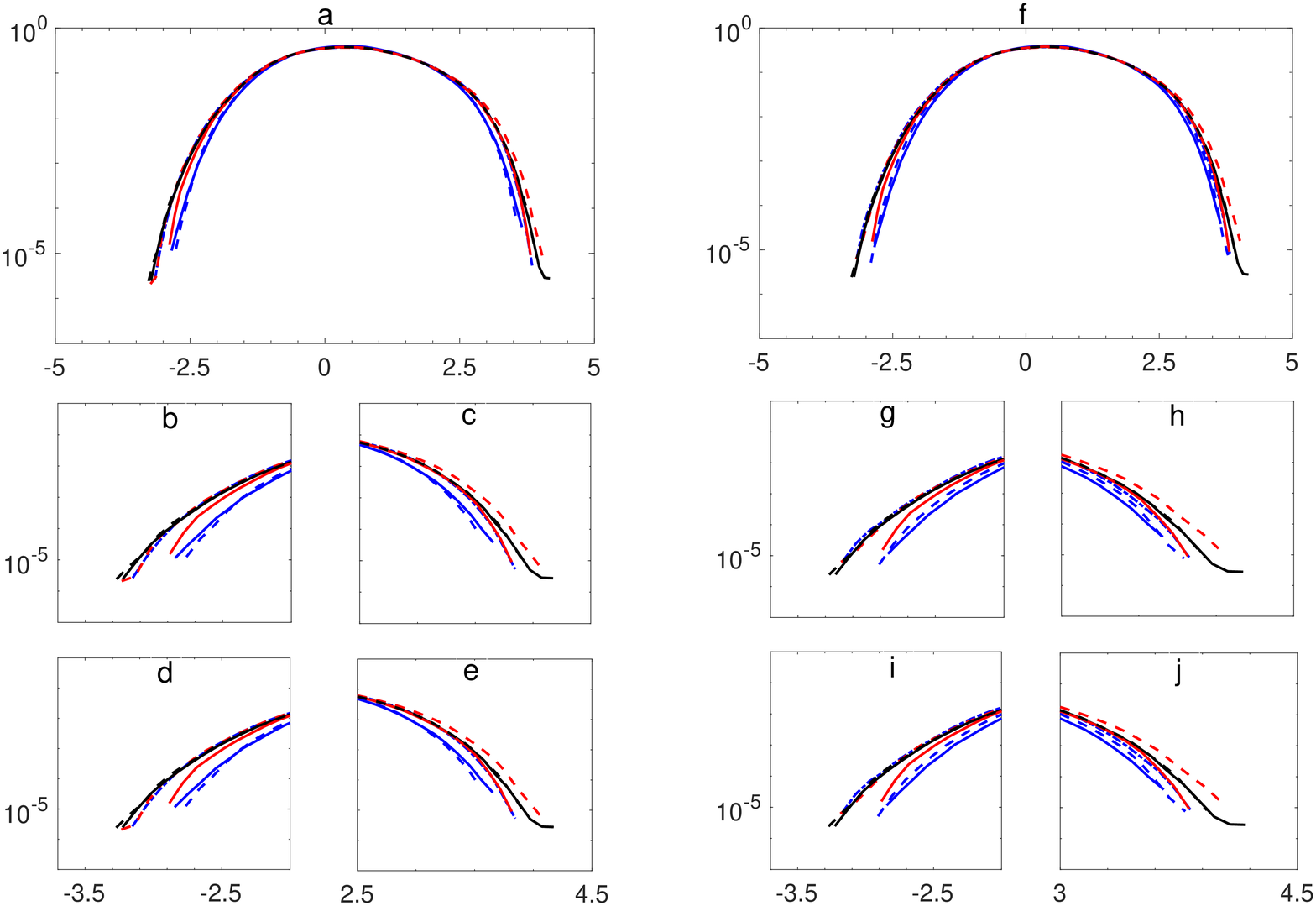}
  \caption{Generalization performance with and without transfer-learning in terms of long-term statistics of Case~2. (a)-(e) Same as Fig.~\ref{pdf} but DNS, HR, SP, and LR are for data from a system with $F=24$ and the ANN or GRU in DD-P, DD-SP, and DD is trained on data from $F=20$. (f)-(j) Same as (a)-(e) except that DD-P, DD-SP, and DD are trained using transfer-learning. Lines show DNS (black solid), HR (black dashed), SP (red solid), DD-SP (red dashed), LR (blue solid), DD-P (blue dashed), and DD (blue dot-dashed).}
\label{TL_pdf}.
\end{figure}

For Case~2 (Fig.~\ref{TL}(b)), both DD-SP and DD fail to generalize. DD-P appears to generalize; however, note that the the accuracy of DD-P for this case (when trained and tested on $F=24$) is lower than that of DD-SP, and in fact after DD-SP's failure to generalize, its performance becomes just comparable to that of DD-P (and LR). The poorer performance of DD-P compared to DD-SP when trained on $F=20$ and tested on $F=24$ can be further seen in the PDFs and their tails (Figs.~\ref{TL_pdf}(a)-(e)). The KS tests (Table~\ref{KS_table_TL}) and the tails show that when trained on $F=20$ and tested on $F=24$ data, DD-P has difficulty with capturing the PDF and its left tail, while DD-SP (and even DD) capture the PDFs and their tails well. This is interesting, as Fig.~\ref{TL}(b) already showed that DD-SP and DD fail to generalize. From what is presented so far, there are a number of lessons to learn

\begin{enumerate}
    \item In evaluating generalization, simply inspecting the PDFs is not enough, as models that might appear to generalize (e.g., DD-SP and DD for Case~2), are in fact not generalizing. Short-term forecasting is a more challenging test for generalization. 
    \item Generalization is much more challenging for fully data-driven models such as DD compared to models such as DD-P and DD-SP, in which the equations of $X$ are still integrated numerically. This is not surprising, as the increase of $F$ does not change anything in the DD model, while it modifies the equations of $X$ in DD-P and DD-SP.
    \item Overall, generalization is a challenge for all models that have a data-driven component. Below, we show how transfer-learning can be used to overcome this challenge.
\end{enumerate}

\begin{figure}[ht]
  \centering
  \includegraphics[width=1\textwidth,scale=0.5]{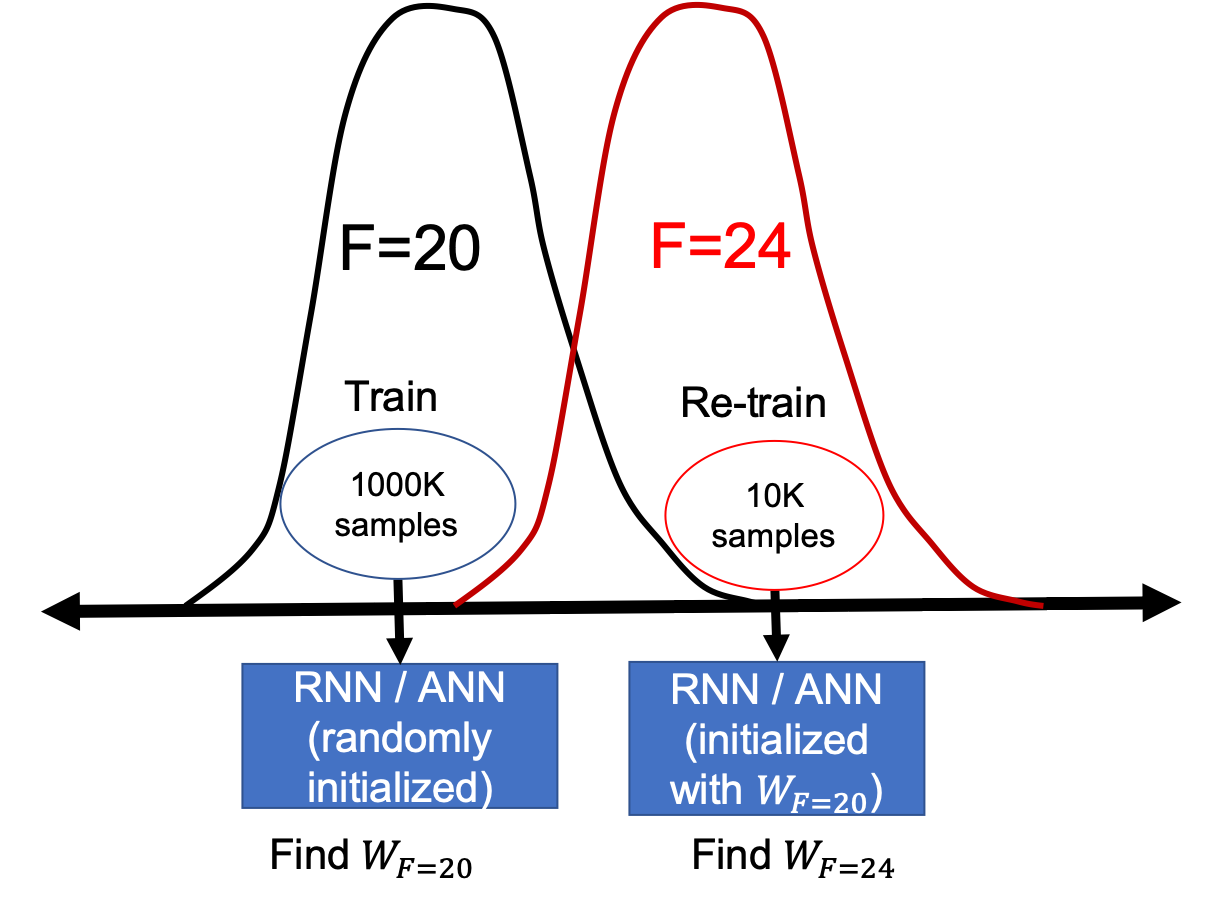}
  \caption{Schematic of the transfer-learning method. $W_{F=20}$ refers to the weights obtained at the end of training for the GRU or ANN on a system with $F=20$ using a large amount of data ($1000$K samples). These weights are used to initialize the GRU or ANN that is re-trained on a much smaller amount of data (just $10$K samples which is $1\%$ of the original training size) from the system with $F=24$.}
\label{TL_schem}.
\end{figure}

So far, for an ANN or GRU, we use a large number of samples ($10^6$) from a system with a given forcing (e.g., $F=20$) and start with randomly initialized weights and train to find weights (e..g., $W_{F=20}$). We show above that $W_{F=20}$ does not generalize to a system with $F=24$. While we can use this approach to find $W_{F=24}$, it would require a large number of samples (e.g., $10^6$) from the system with $F=24$ to obtain accurate results. The idea of transfer-learning \cite{yosinski2014transferable} is that using a well-trained $W_{F=20}$ and a small number of samples (e.g., just $10^4$) from the system with $F=24$, we can find $W_{F=24}$. The idea is shown in schematically in Fig.~\ref{TL_schem}: rather than random initialization of weights to find $W_{F=24}$, the weights are initialized with $W_{F=20}$ and the new $10$K samples are used to re-train the weights. 

Figure~\ref{TL}(a) shows that transfer-learning substantially improves the short-term prediction accuracy of DD for Case~1 and completely closes the generalization gap. Similarly, for Case~2 (Fig.~\ref{TL}(b)), transfer-learning closes the generalization gap for DD-SP and DD. To demonstrate that the improvement in the performance of DD-SP and DD is truly due to the transfer-learning approach, not simply the use of $10$K samples from $F=24$, we also show the results when $10$K samples from $F=24$ are used with randomly initialized weights (no transfer-learning). The DD-P, DD-SP, and DD models trained this way (black asterisks) perform very poorly, indicating the power of transfer-learning when only a small amount of training data from the new system is available. As discussed above, there is no generalization gap in the PDFs of DD-SP or DD (Fig.~\ref{TL_pdf}(a)) and transfer-learning does not affect the PDFs of DD-SP or DD (Fig.~\ref{TL_pdf}(b) and Table~\ref{KS_table_TL}). Figure~\ref{TL_pdf}(b) and the KS tests show that transfer-learning does not improve the PDF of DD-P, and this is because the performance of DD-P is degraded by the lack of scale separation.

\section{Summary and Discussion}
\label{sec:discussion}
In this paper, first we introduce a DD-SP framework, in which the equations of large-scale processes are integrated numerically on a low-resolution grid, and the equations of small-scale processes (which require high-resolution grid to solve) are integrated data-drivenly. The data-driven integration, enabled by the recent advances in deep learning, substantially reduces the computational cost of DD-SP and makes it as cost-effective as parameterized low-resolution models. Figures~\ref{all_cases_noisy} and \ref{pdf} show that DD-SP is greatly superior to LR for both predicting short-term evolution and reproducing long-term statistics, and is superior to deep learning-based DD-P particularly when the system lacks scale separation (Case~2). DD-SP is computationally much cheaper than SP, while it can reproduce the long-term statistics as accurately as SP, and predict the short-term evolution with comparable accuracy as that of SP.  


Second, we examine the ability of DD-P, DD-SP, and DD to generalize to systems that are different from the one they are trained for. By increasing the forcing of the Lorenz system by $20\%$, we find that for Case~1, all models except for the fully data-driven model (DD) generalize, while they have difficulties in generalizing for Case~2. As a remedy, we show that transfer-learning, which involves re-training the deep learning models with a small amount of data from the new system (e.g., $1\%$ of the original dataset size), can substantially improve generalization. 

Note that in the course of this analysis, we find that all models \textit{appear} to generalize for either case even without transfer-learning if, as done commonly, only their PDFs are examined, unless, one closely inspects the tails. However, quantifying the short-term forecasting accuracy shows that some of those models are not, in fact, generalizing. We suggest that to truly evaluate generalization, the performance of models should be examined beyond just simple inspection of PDFs.           


The DD-SP model and transfer-learning both yield promising results in tests using the multi-scale Lorenz 96 system (only for short-term prediction accuracy; transfer learning does not improve the tails of the PDF for long-term statistics). Despite its simplicity, this system allows developing different models that follow the current state-of-the-art modeling approaches (e.g., SP), and clearly reveals the weaknesses of some models in dealing with lack of scale separation and with generalization. However, to assess whether DD-SP and transfer-learning can actually improve the representation of subgrid-scale processes in weather and climate models, more tests using a hierarchy of more complex systems are needed. A two-layer quasi-geostrophic model, a fitting prototype for oceanic \cite{bolton2019applications} and large-scale atmospheric \cite{nabizadeh2019size} turbulent circulations, is a natural next system in this hierarchy.        

In scaling the approaches proposed here up to more complex systems, one potential challenge might be accurate, data-driven integration of complex, high-dimensional governing equations of the small-scale processes. For example, to apply the DD-SP framework to SP-CAM, the 2D cloud-resolving model (CRM) embedded within each point of the low-resolution grid (or at least part of the CRM) should be integrated data-drivenly. Recent studies in the climate science and fluid dynamics communities have shown promising results of fully data-driven 2D or 3D spatio-temporal forecasting of complex, chaotic dynamical systems with RNNs, CNNs, or RNNs+CNNs \cite{dueben2018challenges,scher2018toward,scher2019weather,mohan2019compressed,wang2019towards,weyncan,schergeneralization,chattopadhyay2019analog,chattopadhyay2018test,wu2020enforcing,mohan2020embedding,rasp2020weatherbench}. Such methods, particularly with some key physics constraints enforced, can be used to integrate the 2D or 3D governing equations of the small-scale processes data-drivenly. 

Note that for fully data-driven forecasting of large-scale processes, the accuracy of a the data-driven method (and its generalizability) directly determines the accuracy (and generalizability) of the model, and as discussed in Sections~\ref{sec:shorterm} and ~\ref{sec:gen}, the accuracy and particularly generalizability of such fully data-driven models are limited. However, in DD-SP, the large-scale processes are still integrated numerically, and the inaccuracies of data-driven methods only affect the accuracy of the modeling of the small-scale processes. Said in another word, the numerical solver acts as a \textit{correcting mechanism} on the data-driven integration method. The point is that one may not need very accurate data-driven methods for the DD-SP model to have high overall accuracy because of the use of numerical solver for the large-scale processes.

While the modeling of data-driven small-scale processes is being performed by a potentially less accurate deep learning model, a part of the model's input is coming from more accurate (numerically solved) large-scale processes which act as a \textit{correcting mechanism} for the deep-learning based integration.

Another potential challenge is instabilities arising from coupling of a data-driven model with a numerical model. Our DD-P (with ANN) and DD-SP (with GRU) models are stable (and could be integrated in time for as long as needed); however, such numerical instabilities have been reported for DD-P in more complex systems \cite{o2018using,brenowitz2018prognostic,brenowitz2019spatially,rasp2019online}. The nature of these instabilities and remedies for eliminating/avoiding them should be thoroughly studied when DD-SP and transfer-learning are applied to more complex systems. 

Finally, here we focus on deterministic DD-P and DD-SP. Stochastic parameterization of subgrid processes, for example for dealing with irreducible model uncertainty \cite{palmer2012towards,berner2012systematic,palmer2019stochastic}, can be implemented in ML-based data-driven models, for example using generative adversarial networks (GANs) \cite{gagne2019machine} and physics-constrained GANs \cite{wu2020enforcing}, and should be further pursued in future studies.


\appendix
\section{Details of the ANN used in DD-P}\label{sec:ANN_append}
The ANN used to produce the DD-P results reported in this study has $8$ hidden layers, each of which has $500$ neurons with a $\tanh$ activation function. The input to the ANN has $8$ neurons and takes $X(t)$, a vector of size $8$, as the input, and outputs another vector of size $8$: $\Sigma_j Y_{j,k}$ ($j,k=1,2 \dots 8$). Thus the output is an $8$-neuron layer too. 
The hyperparameters of the ANN, as well as the number of layers, have been optimized with extensive trial and error in the offline mode. Once trained, the ANN provide a data-driven representation for $U_p$ in Eq.~(\ref{eq:LR1}). For both Case 1 and Case 2 we have found the same hyperparameters to yield the best result. The ANN is trained with $10^6$ sequential samples of $(X(t),Y(t))$. 



\section{Details of the GRU used in DD-SP}
\label{append_GRU}
The equations of the GRU \cite{cho2014properties} used here are:
\begin{eqnarray}
  z\left(t\right)=\sigma\left(W^{(z)} x\left(t\right)+ U^{(z)} h\left(t-1\right)\right)
\end{eqnarray}
\begin{eqnarray}
r\left(t\right)=\sigma\left(W^{(r)} x\left(t\right)+ U^{(r)} h\left(t-1\right)\right)
\end{eqnarray}
\begin{eqnarray}
h'(t)=\tanh\left( W x\left(t-1\right) + \left[r(t) \circ U h\left(t-1\right)\right]\right)
\end{eqnarray}  
\begin{eqnarray}
h(t)=\left[z(t) \circ h'(t-1)\right]+\left[z(t-1) \circ h'(t)\right]  
\end{eqnarray}
\begin{eqnarray}
 o(t)=W_{out} h(t) 
\end{eqnarray}
Here, $x \in \Re^{d}$ where $d=72 \times q$ and $q$ is the time delay embedding. The factor $72$ in $d$ comes from the sum of the sizes of $X$ and $Y$: $8$ and $64$. $W^{(z)}$, $W^{(r)}$, $W$, $W_{out}$, $U^{(z)}$, and $U^{(r)}$ are weights to be learnt with the backpropagation-through-time algorithm \cite{goodfellow2016deep}. $z$ is the update state while $r$ is the reset state and $h$ is the hidden state. The operator defined as $\left[a \circ b\right]$ refers to the Hadamard product between $a$ and $b$. 

Hyperparameters are chosen after extensive trial and error for the GRU used in DD-SP. The optimized $q$ is $10$ for Case 1 and $2$ for Case 2. The GRU takes $\left[X(t) \quad Y(t)\right]^T$, $\left[X(t- \Delta t) \quad   Y(t-\Delta t)\right]^T$ $\cdots$ $\left[X(t-(q-1)\Delta t) \quad                Y(t-(q-1)\Delta t)\right]^T$ as input and predicts $\left[X(t+ \Delta t) \quad Y(t+ \Delta t)\right]^T$. The prediction iteratively continues until the GRU predicts $\left[X(t+ 10 \Delta t \quad Y(t+10 \Delta t)\right]^T$ at which point, only $Y(t+ 10 \Delta t)$ is used to feed into the low-resolution RK4 solver for $X$, which uses the previous $X(t)$ and $Y(t+ 10 \Delta t)$ to predict a more accurate $X(t+10 \Delta t)$. This is fed back into the GRU input as it continues predicting for the next time steps. The GRU is trained with $10^6$ sequential samples of $(X(t),Y(t))$. Here, $[a \quad b]^T$ shows concatenation of two vectors $a$ and $b$.

In this work, we choose a GRU as the RNN used for integration of $Y$ in DD-SP. There are other RNNs such as long short-term memory (LSTM) or echo state network (ESN; a.k.a. reservoir computing) that could be used instead (see \citeA{chattopadhyay2019data} for a comparison of LSTM and ESN for DD). However, in this study, we only need short $10
\Delta t$ data-driven integrations in DD-SP, and thus all these methods have equally good performances. Furthermore, in preliminary exploratory experiments, we find using GRU to always lead to stable DD-SP integrations, while using ESN sometimes result in blow ups. We emphasize that we have only done an exploratory set of experiments with ESN; examining the stability of different ML methods in the online mode should be investigated thoroughly in future studies.    

\section{Details of the GRU used in DD}
\label{append_GRU_DD}

For training DD with a GRU, the DNS data has been sampled at every $10 \Delta t$. This has been done in order to keep a fair comparison between all the $X$ solvers' (RK4 for DD-SP, DD-P, and LR integrates at $10 \Delta t$) time step for the parameterized models. The GRU takes in only $X(t), X(t-10 \Delta t), \cdots X(t-10 \Delta t (q-1))$, where $q$ is 3 in both Case 1 and Case 2 and predicts $X(t+10 \Delta t)$.

\acknowledgments
We thank Zhiming Kuang, Krishna Palem, and Devika Subramanian for insightful discussions. Computational resources were provided by NSF XSEDE (allocation ATM170020) to use Stampede2, Bridge GPU, and Comet GPU clusters, and by the Rice University Center for Research Computing. This work was partially supported by an Early-Career Research Fellowship from the Gulf Research Program of the National Academies of Science, Engineering, and Medicine (to P.H.). A.C. thanks the Rice University Ken Kennedy Institute for Information Technology for a BP HPC Graduate Fellowship. All codes for this paper can be found in \url{https://github.com/ashesh6810/Data-driven-super-parametrization-with-deep-learning}. 


%
%

 \bibliography{DDSP_v1}

%
%
%
%
%

\end{document}